# Scaling of Dzyaloshinskii–Moriya interaction at heavy metal and ferromagnetic metal interfaces

Xin Ma[1,*], Guoqiang Yu[2], Chi Tang[3], Xiang Li[2], Congli He[2], Jing Shi[3], Kang L. Wang[2,4], and Xiaoqin Li[1,+]

[1]Department of Physics, The University of Texas at Austin, Austin, Texas 78712, USA
[2]Department of Electrical Engineering, University of California, Los Angeles, California 90095, USA
[3]Department of Physics and Astronomy, University of California, Riverside, California 92521, USA
[4]Department of Physics, University of California, Los Angeles, California 90095, USA

The Dzyaloshinskii–Moriya Interaction (DMI) at the heavy metal (HM) and ferromagnetic metal (FM) interface has been recognized as a key ingredient in spintronic applications. Here we investigate the chemical trend of DMI on the $5d$ band filling ($5d^3 \sim 5d^{10}$) of the HM element in HM/CoFeB/MgO multilayer thin films. DMI is quantitatively evaluated by measuring asymmetric spin wave dispersion using Brillouin light scattering. Sign reversal and 20 times modification of the DMI coefficient $D$ have been measured as the $5d$ HM element is varied. The chemical trend can be qualitatively understood by considering the $5d$ and $3d$ bands alignment at the HM/FM interface and the subsequent orbital hybridization around the Fermi level. Furthermore, a positive correlation is observed between DMI and spin mixing conductance at the HM/FM interfaces. Our results provide new insights into the interfacial DMI for designing future spintronic devices.

The Dzyaloshinskii-Moriya interaction (DMI) refers to a short-range antisymmetric exchange interaction that promotes chiral spin alignments in systems lacking space inversion symmetry [1, 2]. An interfacial DMI can be introduced in the ultrathin ferromagnetic metal (FM) layer adjacent to an anti-ferromagnetic [3] or heavy metal (HM) layer [4] possessing strong spin-orbit (SO) coupling. Such interfacial DMI has received significant attention recently, because its interplay with other SO effects provides a platform for exploring new phenomena promising for spintronic applications [5-10]. For instance, the DMI at FM/HM interfaces is essential to stabilize the Néel type spin configuration in magnetic skyrmions and domain walls with certain chirality [4]. In addition, the direction of skyrmion or domain wall motion driven by an electric current via SO torques is determined by the chirality of the topological spin texture, which in turn, is controlled by the sign of DMI [11-15]. In magnetization switching via SO torques, DMI presents an obstacle since an external magnetic field is required to overcome the chiral domain wall imposed by DMI [16].

The underlying physical principles that determine the DMI at HM/FM interfaces remains unclear despite many previous experimental and theoretical investigations. The DMI coefficient $D$ is estimated on different multilayer structures by previous domain wall studies [17-19], where certain assumptions have to be applied. Direct measurement of $D$ has been recently demonstrated with Brillouin light scattering (BLS) technique [20-30]. However, only a few isolated material systems have been investigated aiming to maximize $D$. So far, no systematic and direct experiment has been reported to investigate the chemical trend of DMI on the choice of material constituents of HM/FM bilayers. In addition, a number of theoretical and experimental studies have investigated the correlation between DMI and other SO effects including SO torques [31-33], proximity induced magnetization [34, 35] and magnetic anisotropy [36]. In light of the important role played by SO interactions in DMI, systematic change of the HM element may be an effective route to seek DMI's correlation with other SO effects and to elucidate the underlying mechanisms of DMI at HM/FM interfaces.

In this letter, we investigate the impact of $5d$ band filling on DMI by systematically changing the $5d$ transition metal layer under the CoFeB/MgO thin film. A wide range variation of the $5d$ band filling ($5d^3 \sim 5d^{10}$) of the HM element leads to significant modification of $D$, which is evaluated quantitatively by measuring the asymmetric spin wave dispersion via Brillouin light scattering (BLS). Sign reversal of $D$ is observed when the $5d$ band occupancy of the HM element changes from less than to more than half filled, similar to the Hund's rule. The strength of $D$ exhibits systematic chemical trend and is maximized with the Pt underlayer, suggesting effective control of DMI by tuning the SO active $5d$ states near the Fermi level. A positive correlation between DMI and spin mixing conductance is observed by changing the HM layer. Such a correlation likely originates from the spin-flip processes between the $3d$ and $5d$ states that impact the DMI strength. Our results may help design material structures with desired $D$ for controlling skyrmions and chiral domain wall dynamics.

A series of X(5)/Co$_{20}$Fe$_{60}$B$_{20}$(1)/MgO(2)/Ta(2) (X = Ta, W, Ir, Pt) thin films were deposited by magnetron sputtering at room temperature on thermally oxidized silicon substrates, where the numbers in parentheses denote the nominal layer thicknesses in nanometers. Co$_{20}$Fe$_{60}$B$_{20}$(1)/ MgO(2)/Ta(2) multilayers were also sputtered onto Au(5) underlayer prepared with e-beam evaporation. No further annealing procedure was applied after the deposition, in order to minimize the inter-diffusion of atoms between different layers [37]. As a result, the CoFeB and HM layers are amorphous or polycrystalline. For those structures with X = W, Ir, and Pt, 5 nm-thick Ta seed layer was placed at the bottom of X HM underlayer to improve the HM/CoFeB interface quality [27]. The HM layer thickness is chosen as 5 nm to saturate the DMI strength, excluding the variance of DMI introduced by HM thickness [23, 24]. We chose CoFeB as the candidate for the FM layer, because perpendicular magnetic anisotropy (PMA) can be easily established at the CoFeB/MgO interface [24], which is desirable for room temperature skyrmions [38] and enhances the tunnel magneto-resistance ratio in magnetic tunneling junctions [39]. These features may make our results relevant for existing spintronic devices.

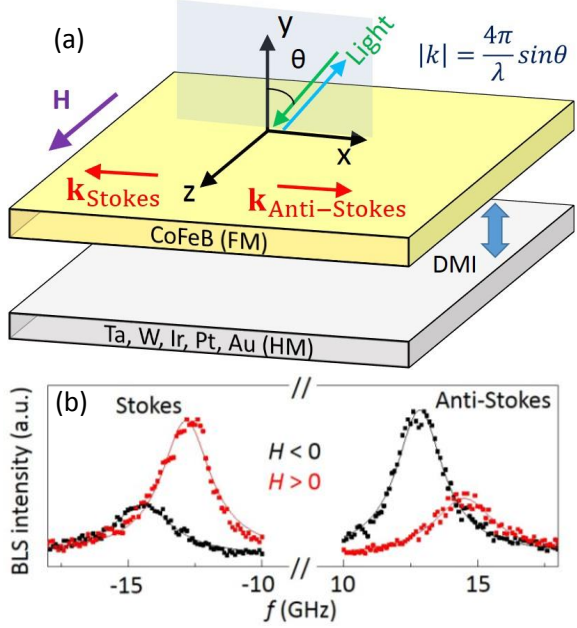

Fig. 1. (a) Schematics of BLS measurement geometry. (b) BLS spectra for DE spin waves recorded at a fixed incident angle $k = 16.4$ rad/μm under oppositely oriented external magnetic fields **H** in Pt/CoFeB/MgO. The solid lines represent fittings with Lorentzian functions.

BLS measurements were performed to investigate the frequency difference between counter-propagating Damon-Eshbach (DE) spin waves induced by DMI [3, 24]. Figure 1a shows the geometry of the BLS experiment, where an in-plane magnetic field **H** was applied along the $z$ axis in all measurements. An s-polarized laser beam was incident on the sample, and the p-polarized component of the backscattered light was collected and sent to a Sandercock-type multipass tandem Fabry-Perot interferometer. In the light scattering process, the total momentum is conserved in the plane of the thin film. As a result, the Stokes (anti-Stokes) peaks in BLS spectra correspond to the creation (annihilation) of magnons with momentum $|k| = \frac{4\pi}{\lambda} \sin\theta$ along $-x$ ($+x$) direction as illustrated in Fig. 1a, where $\lambda = 532\ nm$ is the laser wavelength, and $\theta$ refers to the light incident angle.

Figure 1b displays typical BLS spectra for DE spin wave modes under opposite **H** directions on the Pt/CoFeB/MgO thin film. The frequencies of the Stokes ($-k$) and anti-Stokes ($k$) peaks are different, while the frequencies corresponding to $\pm k$ can be interchanged with each other by reversing the **H** direction. The spin wave frequency can be described by [3]

$$f = \frac{\gamma}{2\pi}\sqrt{\left(H + \frac{2A}{M_S}k^2 + 4\pi M_S(1 - \xi(kt)) - \frac{2K_\perp}{M_S}\right) *}$$
$$\sqrt{\left(H + \frac{2A}{M_S}k^2 + 4\pi M_S \xi(kt)\right)} - \varepsilon(K_\perp, k * sgn(M_z)) -$$
$$sgn(M_z)\frac{\gamma}{\pi M_S}Dk \tag{1}$$

where $H$ is the external field, $\gamma$ is the gyromagnetic ratio, $A$ is the exchange stiffness constant, $M_S$ is the saturated magnetization, $\xi(kt) = 1 - (1 - e^{-|kt|})/|kt|$ with $t$ being the CoFeB thickness, $K_\perp$ is the interfacial magnetic anisotropy which mainly originates from the CoFeB/MgO interface [25], and $\varepsilon(K_\perp, k)$ describes a correction in frequency due to interfacial anisotropy and non-reciprocity as discussed below. Both $D$ and $k$ can be positive or negative values in the formula. In Eq.1, the first term even in $k$ on the right hand side describes the spin wave dispersion without DMI. The second term is much smaller than the DMI effect as discussed below. We take into account this second term explicitly in all analyses of DMI. Most importantly, the third term accounts for the frequency difference between counter-propagating spin waves induced by DMI and is odd in $k$.

To quantify the interfacial DMI constant $D$, momentum ($k$) resolved BLS measurements were performed through varying the incident angle. According to Eq. 1, we can simplify the determination of $D$ by subtracting the BLS spectra.

$$\Delta f = \frac{\left((f(-k,M_z) - f(k,M_z)) - (f(-k,-M_z) - f(k,-M_z))\right)}{2}$$
$$= \frac{2\gamma}{\pi M_S}Dk + \Delta\varepsilon(k) \tag{2}$$

where $\Delta\varepsilon(k) = \varepsilon(K_\perp, k) - \varepsilon(K_\perp, -k)$ is much smaller than the DMI term. In order to determine $D$ accurately, we first estimated $\Delta\varepsilon(k)$ on different samples. $\Delta\varepsilon(k)$ originates from the non-reciprocity of DE spin waves in the presence of interfacial magnetic anisotropy $K_\perp$. In the DE geometry depicted in Fig.1a, spin waves propagating to $-x$ $(+x)$ direction localize near top CoFeB/MgO (bottom HM/CoFeB) interface, experience stronger (weaker) perpendicular anisotropy field $2K_\perp/M_S$ from CoFeB/MgO interface, and hence undergo a decrease (an increase) in the spin wave frequencies relatively. We first determine the $K_\perp$ values through the $H$ dependence of spin wave frequency with $k = 0$ using normal incident light and fitting with Eq. 1, as shown in Fig. 2a. Then, we use a mean-field approach to estimate $\Delta\varepsilon(k)$ (see supplemental material for details [40]). Figure 2b plots the dependence of $\Delta\varepsilon(k)$ on $k$ for different samples. The sample with Ta underlayer exhibits relative larger $|\Delta\varepsilon(k)|$, because of its strong perpendicular magnetic anisotropy. After correcting $\Delta f$ with $\Delta\varepsilon(k)$ in Eq.2, the slopes of the linear correlations can be used to determine the $D$ values [41].

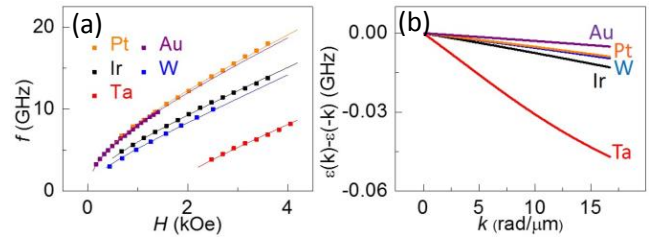

Fig. 2. (a) The dependence of spin wave frequency $f(k = 0)$ on $H$ in CoFeB samples with different underlayers. The solid lines are fittings with Eq.1 under $k = 0$. (b) Simulation results of $\varepsilon(K_\perp, k) - \varepsilon(K_\perp, -k)$.

Figures 3a, b plot the measured frequency difference $\Delta f$ as a function of $k$ for the X/CoFeB/MgO (X = Ta, W, Ir, Pt, Au, MgO, and Pt/Cu(1)) multilayer thin films. The data is well fitted by a linear function in all cases as described in Eq.2. Different slopes of the linear fittings on these samples mainly results from the change of $D$, because the magnitude of $\Delta f$ is much larger than the $\Delta\varepsilon(k)$ in Fig. 2b. DMI is absent in the MgO/CoFeB/MgO thin film due to the recovery of inversion symmetry. Moreover, DMI strength drops significantly by inserting 1 nm Cu in between Pt and CoFeB layers [42]. No significant spin relaxation is expected in transversing the 1 nm Cu spacer between the Pt and CoFeB layers. The drastically reduced DMI results from disrupted hybridization between the $3d$ (CoFe) and $5d$ (Pt) orbitals as we elaborate below.

Our key finding in the chemical trend of DMI on $5d$ band filling of the HM element is summarized in Fig. 3c. Qualitative agreement between our results and previous first principle calculations [30, 34, 43] has been found with respect to the general chemical trend of DMI sign and magnitude, as discussed below. Quantitative agreement between theory and experiment is difficult to achieve, because a certain crystalline structure has to be assumed for HM and FM layers in theoretical models, differing from the amorphous or polycrystalline material structure used in our study and those in practical spintronic devices.

Evidently, the interfacial DMI reverses sign when the HM moves towards heavier element in Fig. 3c. The samples with Ta, W, and Ir underlayers exhibit $D > 0$ with right-handed magnetic chirality preferred, while those with Pt and Au underlayers have $D < 0$ with left-handed magnetic chirality preferred. The sign of DMI is related to the $5d$ electron filling in the HM. Among the samples we investigated, the $5d$ bands of Ta and W are less than half filled by electrons, while those of Pt and Au are more than half filled. Such difference in $5d$ electron occupancy leads to opposite signs in the expectation value of SO coupling $<\mathbf{l} \cdot \mathbf{s}>$ between Ta (W) and Pt (Au) according to the Hund's rule. Originating from the SO coupling in the HM, the DMI sign should also be reversed between samples on Ta (W) and Pt (Au) underlayers, consistent with our observation in Fig. 3c. This interpretation agrees with a recent theoretical study that highlighted the important role of the Hund's rule in determining DMI at HM/FM interfaces [43]. We caution that the DMI sign for the sample with Ir underlayer seems an outlier and is still under debate in the literature. DMI constants with opposite signs have been reported experimentally at various Ir/FM interfaces of multilayer thin films (FM = Co, Ni, (Ni/Co)$_N$, and CoFeB) [3, 19, 26, 44], and theoretical calculations also show that DMI changes sign between Ir/Co and Ir/Fe [43]. Because the $5d$ electron filling of Ir sits near the transition point of DMI sign reversal, the DMI sign becomes very sensitive to the $3d$ band alignment of the FM in Ir/FM multilayers.

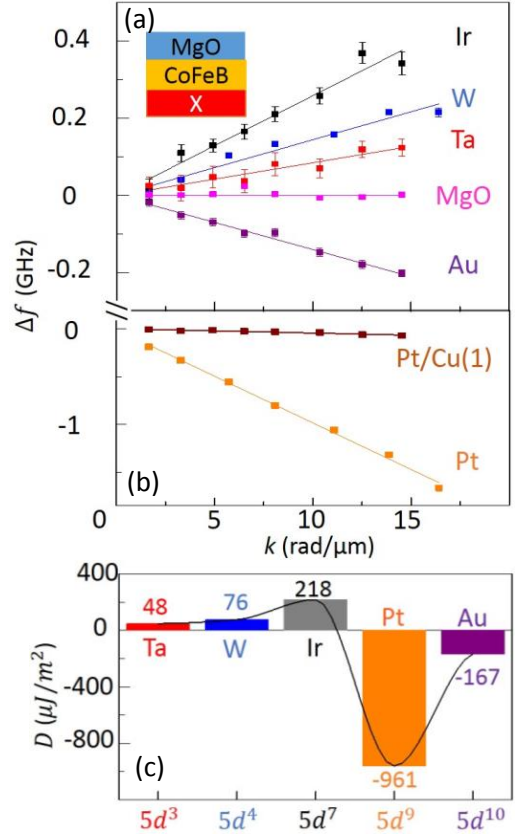

Fig. 3. (a, b) The linear dependence of $\Delta f$ on $k$ in CoFeB samples with different underlayers. (c) The $D$ values determined at CoFeB samples with different underlayers. The black line serves as a guide for the eye.

The magnitude of DMI maximizes with Pt underlayer and decreases as the number of $5d$ electrons either increases or decreases in the HM element. In addition, there is a dramatic change (~ 20 times) in DMI strength between Ta and Pt and a sudden drop of the DMI strength from Pt to Au. This trend can be qualitatively understood by considering the $5d$ and $3d$ bands alignment at the HM/FM interface and the subsequent orbital hybridization around the Fermi level [43]. The increase of DMI strength from Ta to Pt is owing to the relocation of Fermi level with respect to the $5d$ band of the HM. At different location (energy) of the broad $5d$ band, the $5d$ states are mainly with different orbital characters (i.e. $d_{xy}, d_{yz}, d_{xz}, d_{z^2}, d_{x^2-y^2}$) and hence contribute differently to DMI due to their varied degree of hybridization with $3d$ states [30]. By increasing the $5d$ electron number (i.e. Ta $5d^3$ to Pt $5d^9$), the Fermi level relocates towards the $5d$ states with certain orbital characters. These $5d$ states around Fermi level facilitate the spin-flip transitions between occupied and unoccupied $3d$ states, and hence selectively dominate the overall DMI [43]. The above qualitative arguments are supported by a first-principle calculation performed on a simple model system consisting of $5d$-$3d$ transition metal zigzag chains [30], where $5d_{xz}$ and $5d_{yz}$ states yield stronger contribution to the DMI. Moving these states towards the Fermi level by incorporating $5d^7 \sim 5d^9$ HM elements leads to larger DMI. This calculation predicts a maximal DMI strength associated with Pt among all HM materials investigated and is consistent with our observations. Similarly, the sharp drop of DMI strength observed with Au layer originates from the absence of $5d$ states at the Fermi level and the subsequent reduction of $5d$-$3d$ orbital hybridization [34].

Finally, we report an observed correlation between DMI and spin pumping effect at HM/FM interfaces. As discussed above, the enhancement of DMI by modifying $5d$-$3d$ hybridization is through facilitating the spin-flip transitions between $3d$ states that also involve transitions with $5d$ SO active states (i.e. $3d$-$5d$-$3d$ electron hopping). Such processes also increase the effective spin mixing conductance $g_{eff}^{\uparrow\downarrow}$ in the spin pumping effect [45]. Therefore, one may expect a positive correlation between $D$ and $g_{eff}^{\uparrow\downarrow}$ by changing the HM underlayer. The spin mixing conductance $g_{eff}^{\uparrow\downarrow}$ is determined through the spin pumping enhanced damping $\alpha_{sp}$ extracted via the full width half maximum (FWHM) of the BLS spetra [26] [40]. Figures 4a-d present the BLS linewidth FWHM as a function of $H$, which can be well fitted with FWHM $= \frac{2\alpha\gamma}{\pi}H + \delta f_0$. Here, $\delta f_0$ is the extrinsic linewidth unrelated to $H$ [46], and $\alpha = \alpha_{sp} + \alpha_0$. The intrinsic damping of CoFeB layer $\alpha_0$ is estimated by measuring a sample with MgO underlayer in which the spin pumping is assumed to be absent or $\alpha_{sp} = 0$ in Fig. 4d. The slopes of the linear dependence between FWHM and $H$ as shown in Figs. 4a-c are used to determine $\alpha_{sp}$. Furthermore, we determine spin mixing conductance using $g_{eff}^{\uparrow\downarrow} = \frac{4\pi M_s t_{\text{CoFeB}}}{\gamma \hbar}\alpha_{sp}$ at different HM/FM interfaces.

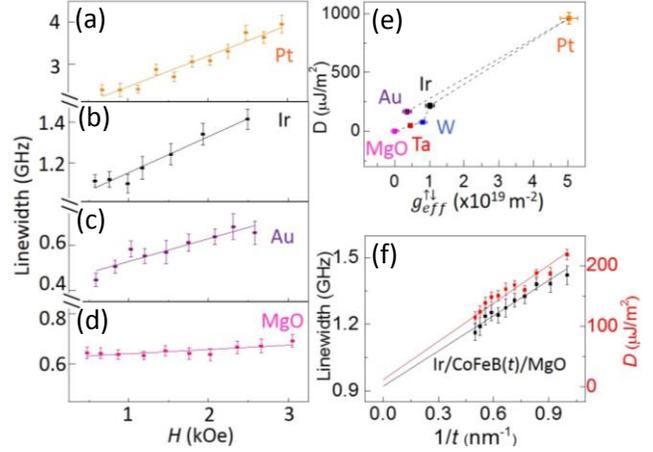

Fig. 4. (a-d) The dependence of spin wave resonance linewidth FWHM on $H$ in CoFeB samples with different underlayers. (e) A positive correlation between $D$ and $g_{eff}^{\uparrow\downarrow}$. The dashed lines guide the trend to heavier element. (f) The $D$ (red) and linewidth FWHM (black) as a function of $1/t$ (CoFeB) at Ir/CoFeB(wedge)/MgO sample. The solid lines are the least square fits.

The estimated $\alpha_{sp}$ and $g_{eff}^{\uparrow\downarrow}$ on different samples agree well with the values found in the literature [47-52][53]. Figure 4e shows a positive correlation between $D$ and $g_{eff}^{\uparrow\downarrow}$ by changing the HM layer at HM/FM interfaces, consistent with the interpretation that DMI is driven by the spin-flip processes between $3d$ and $5d$ states [54]. Finally, we report that both $D$ and FWHM are inversely proportional to the CoFeB thickness $t$ (Fig. 4f), confirming that both the DMI and spin pumping effect originate from the HM/FM interface.

In conclusion, we investigate the dependence of the interfacial DMI on the $5d$ transition metal underlayer at HM/CoFeB/MgO multilayer thin films. The DMI coefficient changes by an order of magnitude and reverses sign when the HM moves towards heavier element in the $5d$ transition metal. The observations can be mostly explained by distinct degree of hybridization between $3d$ and $5d$ orbitals near the Fermi level owing to different $5d$ electron filling in the HM element. A correlation between DMI and spin mixing conductance, two interfacial effects at HM/FM interface, is observed, indicating the important role of spin-flip mixing processes in DMI. We anticipate that our results will provide guidance for designing magnetic structures with desired $D$ and chiral properties in ultra-thin magnetic films.

We gratefully acknowledge helpful discussions with Allan MacDonald, Pantelis Lapas, and Hua Chen. The collaboration among UT-Austin, UCLA and UCR are supported by SHINES, an Energy Frontier Research Center funded by the U.S. Department of Energy (DoE), Office of Science, Basic Energy Science (BES) under award # DE-SC0012670.

Emails: *xma518@utexas.edu
        +elaineli@physics.utexas.edu

[54] Fluctuations in the correlation are anticipated, because the spin mixing conductance depends on other physical parameters or mechanisms which may vary by changing the HM layer.